\pdfoutput=1

\documentclass{article}
\usepackage[utf8]{inputenc}
\usepackage{spconf}
\usepackage{graphicx}
\usepackage{amsmath}
\usepackage{amssymb}
\usepackage{amsfonts}
\usepackage{amsthm}
\usepackage{float}
\usepackage{siunitx}
\usepackage{enumitem}
\usepackage{placeins}
\usepackage{booktabs}
\usepackage{tabularx}
\usepackage{bm}
\usepackage{comment}
\usepackage{mathtools}
\usepackage{xcolor}
\usepackage{lipsum}

\usepackage{hyperref}
\usepackage{cleveref}

\crefname{figure}{Fig.\@}{Fig.\@}

\title{Autonomous Soundscape Augmentation\\with Multimodal Fusion of Visual and Participant-linked Inputs}
\name{
    Kenneth Ooi$^1$,
    Karn N. Watcharasupat$^2$,
    Bhan Lam$^1$,
    Zhen-Ting Ong$^1$,
    Woon-Seng Gan$^1$
\thanks{This research is supported by the Singapore Ministry of National Development and the National Research Foundation,  Prime Minister's Office under the Cities of Tomorrow Research Programme (Award No.\@ COT-V4-2020-1). 
Any opinions, findings and conclusions or recommendations expressed in this material are those of the authors and do not reflect views of the National Research Foundation and Ministry of National Development, Singapore.
}
}
\address{
    $^1$School of Electrical and Electronic Engineering, Nanyang Technological University, Singapore\\
    $^2$Center for Music Technology, Georgia Institute of Technology, Atlanta, GA, USA\\ % 
    Emails: wooi002@e.ntu.edu.sg, kwatcharasupat@gatech.edu, \{bhanlam, ztong, ewsgan\}@ntu.edu.sg
}

\usepackage{fancyhdr}

\fancypagestyle{firstpage}{
     \fancyhf{}
     \fancyhead[L]{Submitted to the 2023 IEEE International Conference on Acoustics, Speech and Signal Processing.}
     \fancyfoot[L]{\footnotesize\textcopyright 2023 IEEE. Personal use of this material is permitted. Permission from IEEE must be obtained for all other uses, in any current or future media, including reprinting/republishing this material for advertising or promotional purposes, creating new collective works, for resale or redistribution to servers or lists, or reuse of any copyrighted component of this work in other works.}
}
 
\begin{document}
\ninept
\maketitle

\begin{abstract}
Autonomous soundscape augmentation systems typically use trained models to pick optimal maskers to effect a desired perceptual change. While acoustic information is paramount to such systems, contextual information, including participant demographics and the visual environment, also influences acoustic perception. Hence, we propose modular modifications to an existing attention-based deep neural network, to allow early, mid-level, and late feature fusion of participant-linked, visual, and acoustic features. Ablation studies on module configurations and corresponding fusion methods using the ARAUS dataset show that contextual features  improve the model performance in a statistically significant manner on the \textit{normalized ISO Pleasantness}, to a mean squared error of $\num{0.1194}\pm\num{0.0012}$ for the best-performing all-modality model, against $\num{0.1217}\pm\num{0.0009}$ for the audio-only model. Soundscape augmentation systems can thereby leverage multimodal inputs for improved performance. We also investigate the impact of individual participant-linked factors using trained models to illustrate improvements in model explainability. % ± 
\end{abstract}


\section{Introduction}\label{sec:Introduction}

The soundscape approach to noise control, as defined in ISO 12913, recommends assessments of the ``acoustic environment as perceived or experienced and/or understood by a person or people, in context'' \cite{InternationalOrganizationforStandardization2014}. Accordingly, soundscape practitioners often focus on interventions that alter the \textit{perception} of acoustic environments, mindful that simply reducing the sound pressure level of a noisy environment may not correlate well with improved perception \cite{DePaivaVianna2015NoiseStudy, Kang2019}. One such intervention is \textit{soundscape augmentation}, which introduces ``maskers'' as additional sounds via electroacoustic systems to improve the perception of the acoustic environment. 
The choice of maskers can be done manually, in an expert-driven \cite{Coensel2011} or participant-driven fashion~\cite{VanRenterghem2020}, but autonomous systems, such as those described in \cite{Jahani2021AnAreas} and \cite{Wong2022DeploymentAugmentation}, can reduce the time and labor required in manual approaches.

However, a weakness of existing autonomous systems is the absence of consideration for the \textit{context} of perception, which is crucial for modeling perceptual attributes, such as the perceived pleasantness of soundscapes \cite{Mitchell2021b}. Contextual factors known to affect soundscape assessments include listener- or participant-linked demographic variables, such as age \cite{Yang2005AcousticSpaces}, responses to self-reported psychological questionnaires \cite{Aletta2018, Ratcliffe2021}, and the present activity while experiencing the soundscape \cite{Mitchell2020TheInformation}. In addition, factors related to the visual environment, such as the physical objects present in a scene \cite{Preis2015}, or the proportion of landscape elements like greenery and buildings \cite{PuyanaRomero2016, Tan2022TheCity},  may also affect soundscape assessments.

Hence, this study aims to improve the performance of a model deployable in an autonomous soundscape augmentation system, by additionally fusing visual and participant-linked information to the existing acoustic information captured by the model. We use an attention-based deep neural network (DNN) architecture previously designed for a purely-acoustic prediction model of perceptual soundscape attributes \cite{Watcharasupat2022AutonomousGain}, and propose new approaches for the DNN to \textit{optionally} exploit the visual and participant-linked information when available or desired. These approaches naturally extend the functions of existing model components performing feature augmentation and probabilistic output prediction, while having a threefold advantage over the pre-existing model: (a) the modified models are backward-compatible with the pre-existing audio-only version as detailed in \Cref{sec:Proposed Method}, (b) their use of additional modalities can improve model performance in mean squared error (MSE) of predictions based on our validation experiments described in \Cref{sec:Validation Experiments}, and (c) they can be used to explain perceptual differences based on participant-linked factors as illustrated in \Cref{sec:Results and Discussion}. 

\vspace{-1.5mm}

\section{Related Work}\label{sec:Related Work}

Multimodal models for a given task usually use inputs corresponding to different types of sensors. Raw data from these sensors can potentially have differing representations, so a possible way to handle this is to train a sub-network for each input modality and aggregate either intermediate features or predictions extracted from each sub-network \cite{Baltrusaitis2019MultimodalTaxonomy}.
For example, in an audio-visual scene classification task, \cite{Okazaki2021AVariants}~combined a visual sub-network of CLIP encoders, which were pre-trained with contrastive learning on image and textual data, with a convolutional sub-network for input audio, and concatenated intermediate features from the audio and visual sub-networks before making a final prediction of the location where a given recording was made. Similarly,
\cite{Naranjo-Alcazar2021Squeeze-ExcitationClassification}~combined a pre-trained VGG16 sub-network for input video with a convolutional sub-network for input audio, but used predictions from both sub-networks in an ensemble classifier. Both methods had improved accuracy when using all modalities as opposed to using any one individual modality. This demonstrates the potential synergy of multimodal inputs, since one modality can supplement information missing from another, thereby allowing multimodal models to capitalize on all available information~\cite{Baltrusaitis2019MultimodalTaxonomy}.

In a similar manner, neural attention-based mechanisms have also been popular as a multimodal feature alignment technique. Such mechanisms aim to mimic the human capability to focus on pertinent data, by assigning weights to features obtained from the different modalities denoting their relative importance. For example, \cite{Priyasad2020AttentionRecognition} explored the use of self-attention mechanisms for speaker emotion recognition on networks taking inputs from the textual and acoustic modalities, and found that applying self-attention before the fusion of multimodal features slightly increased classification accuracy, as compared to doing so after fusion. Similarly, \cite{Ma2019AttnSense:Recognition} used self-attention for human activity recognition, where the attention weights were distributed across features extracted from multiple accelerometers and gyroscope sensors by a convolutional recurrent neural network.

Research on multimodal perceptual models for soundscapes has primarily been centered on the use of hand-crafted features from the acoustic and visual modalities in linear regression models and shallow neural networks \cite{Lionello2020ASoundscapes}, presumably due to the ease of implementation and straightforward explainability. Nonetheless, neural attention has a natural parallel in soundscape perception, via the concept of auditory salience of acoustic events \cite{Huang2017AuditorySoundscapes}. In \cite{Watcharasupat2022AutonomousGain}, a framework for a probabilistic perceptual attribute predictor (PPAP) carrying out soundscape augmentation in the feature domain was proposed, where the ``probabilistic'' loss function
\begin{align}
   	\mathcal{J} &= K^{-1}\sum_{k}\left[\left((y_k-\widehat{\mu}_k)/{\widehat{\sigma}_k}\right)^2/2+\log\widehat{\sigma}_k\right],
    \label{eq:probabilistic_loss}
\end{align}
\vspace{-3mm}

\noindent defined in \cite{Ooi2022ProbablyAugmentation},
was used to train an attention-based DNN to account for inherent randomness in perceptual responses for a batch of $K$ samples, with the model predicting the distribution of the $k$-th response as $\mathcal{N}(\widehat{\mu}_k,\widehat{\sigma}_k^2)$, and $y_k$ being a ground-truth \textit{observation} of the $k$-th response. This is equivalent to the negative log-probability of observing $y_k$, if its ``true'' distribution were $\mathcal{N}(\widehat{\mu}_k,\widehat{\sigma}_k^2)$.

However, the existing framework, which we term the ``audio-only PPAP'' (\textsc{aPPAP}), uses only acoustic features to make predictions, without information that may affect perceptual responses like the visual environment and participant-linked parameters. We thus propose to modify the \textsc{aPPAP} to include features extracted from the visual environment and participant context as additional conditioning inputs, thereby transforming it into a ``contextual PPAP'' (\textsc{cPPAP}) that learns from multiple modalities.

\section{Proposed Method}\label{sec:Proposed Method}

\subsection{Audio-only PPAP (\textsc{aPPAP})}\label{sec:Proposed Method/Audio-only PPAP}

Consider the log-mel spectrogram of a soundscape $\bm{s}\in\mathbb{R}^{T\times F\times C_{\text{s}}}$ with $T$ time bins, $F$ mel bins, and $C_{\text{s}}$ channels, and the log-mel spectrogram of a (possibly silent) masker $\bm{m}\in\mathbb{R}^{T\times F\times 1}$ used to augment $\bm{s}$. In practice, $\bm{m}$ is reproduced by adjusting its digital gain via a multiplicative factor of $10^{\gamma}$, where $\gamma$ is the masker log-gain, before actual playback. The task of the \textsc{aPPAP} is to predict the values $\mu$ and $\log\sigma$, characterizing the distribution $\mathcal{N}(\mu,\sigma^2)$ of some perceptual attribute of interest (e.g., \textit{pleasantness} or \textit{eventfulness}), when $\bm{s}$ is augmented with $\bm{m}$ at a digital gain of $10^{\gamma}$.

To do so, the \textsc{aPPAP} extracts relevant soundscape embeddings 
$\bm{k}=f_{\text{s}}(\bm{s})\in\mathbb{R}^{N\times D}$ and masker embeddings $\bm{q} = f_{\text{m}}(\bm{m})\in\mathbb{R}^{N\times D}$, where $f_{\text{s}}$ and $f_{\text{m}}$ are the respective feature extractors, $N$ is the number of compressed time frames, and $D$ is the embedding dimension. The embeddings are then combined with the masker log-gain $\gamma$ in a feature augmentation block $f_{\text{g}}$ to obtain the augmented soundscape embeddings $\bm{v} = f_{\text{g}}(\bm{k},\bm{q},\gamma)\in\mathbb{R}^{N\times D}$. The masker embeddings $\bm{q}$ are then used to query a mapping from the soundscape embedding ``keys'' $\bm{k}$ to the augmented soundscape embedding ``values'' $\bm{v}$ via a QKV attention block $f_{\text{a}}$, which returns a set of embeddings $\bm{z} = f_{\text{a}}(\bm{q},\bm{k},\bm{v})\in\mathbb{R}^D$. The embeddings $\bm{z}$ are finally used in the output block $f_{\text{o}}$ to predict $\mu$ as $\widehat{\mu}$ and $\log\sigma$ as $\log\widehat{\sigma}$. The logarithms of the standard deviation $\sigma$ and multiplicative factor $10^{\gamma}$ are used instead of their actual values for numerical stability.

\subsection{Contextual PPAP (\textsc{cPPAP})}\label{sec:Proposed Method/Contextual PPAP}

\begin{figure}
    \centering
    \includegraphics[width=\columnwidth]{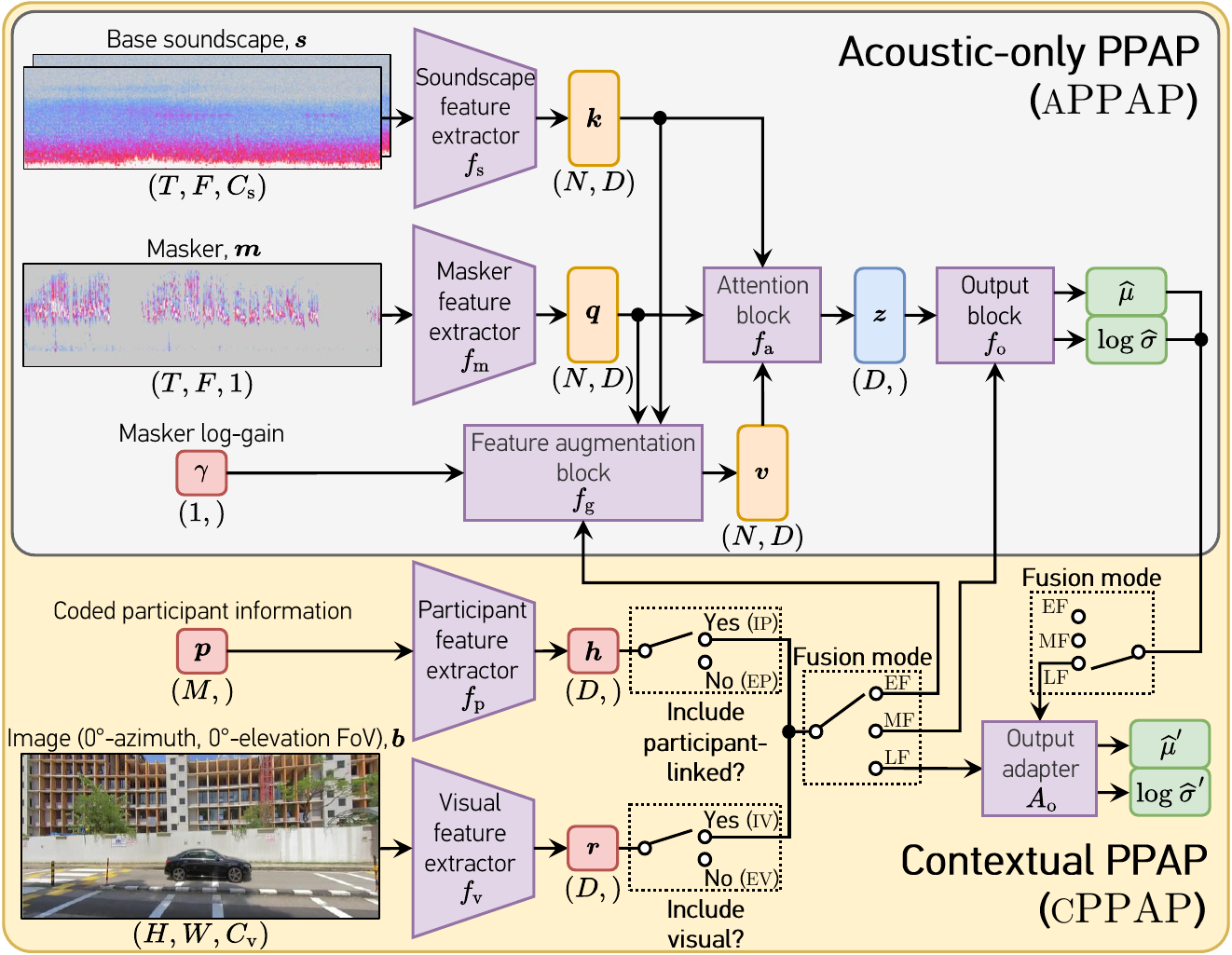}
    \caption{Architecture of audio-only and contextual PPAP. Switches indicate the different configurations of the PPAP used for our validation experiments. Abbreviations: \textsc{ip}/\textsc{ep} = include/exclude $\boldsymbol{h}$; \textsc{iv}/\textsc{ev} = include/exclude $\boldsymbol{r}$; \textsc{ef}/\textsc{mf}/\textsc{lf} = early/mid-level/late fusion.}
    \vspace{-1mm}
    \label{fig:demo-vis-all}
\end{figure}

The contextual PPAP (\textsc{cPPAP}) is first modified from the \textsc{aPPAP} by incorporating inputs from two other modalities:
\begin{enumerate}[wide=0pt, widest=99, leftmargin=\parindent, labelsep=*]
    \item \textbf{Participant modality:} A vector of coded participant information $\bm{p}\in\mathbb{R}^{M}$, where $M$ is the number of participant-linked features used as input. This could contain any numerical representation of information associated with a real or hypothetical participant experiencing the soundscape $\bm{s}$.
    \item \textbf{Visual modality:} A static image $\bm{b}\in\mathbb{R}^{H\times W \times C_{\text{v}}}$, where $H$, $W$, $C_{\text{v}}$ are respectively the height, width, and number of color channels. This could correspond to a picture of the \textit{in-situ} environment where $\bm{s}$ is experienced by the participant.
\end{enumerate}
The task of the \textsc{cPPAP} is similarly to predict the values $\mu$ and $\log\sigma$ characterizing the distribution $\mathcal{N}(\mu,\sigma^2)$ of the same perceptual attribute of interest when $\bm{s}$ is augmented with $\bm{m}$ at a digital gain of $10^{\gamma}$, but with additional information about the person rating that perceptual attribute in $\bm{p}$ and the physical location where the soundscape augmentation occurs in $\bm{b}$. A visual representation of the audio-only and contextual PPAPs is shown in \Cref{fig:demo-vis-all}.

To utilize the information in $\bm{p}$ and $\bm{b}$, we propose to extract relevant participant embeddings $\bm{h} = f_{\text{p}}(\bm{p})\in\mathbb{R}^{D}$ and visual embeddings $\bm{r} = f_{\text{v}}(\bm{b})\in\mathbb{R}^{D}$ using feature extractors $f_{\text{p}}$ and $f_{\text{v}}$. The embeddings $\bm{h}$ and $\bm{r}$ can then be incorporated into the \textsc{aPPAP} to modify it into the \textsc{cPPAP} via early fusion (\textsc{ef}), mid-level fusion (\textsc{mf}), or late fusion (\textsc{lf}), which we individually investigate for our validation experiments, and explicitly define in the following subsections. The  fusion methods in the \textsc{cPPAP} are designed to preserve the modularity of the \textsc{aPPAP}, such that information from any non-acoustic modality can be omitted by zeroing out the embeddings $\bm{h}$ and $\bm{r}$ at any fusion stage. This could be useful, for example, at inference time when information from specific modalities is unavailable due to unforeseen real-life deployment conditions.

\subsubsection{Early fusion \textup{(\textsc{ef})}}\label{sec:Proposed Method/Contextual PPAP/Early fusion}

In \textsc{ef}, the feature augmentation block $f_{\text{g}}$ is modified to fuse information from all modalities such that they are included in the augmented soundscape embeddings $\bm{v}$.
All modalities are thus jointly represented in $\bm{z}$.
We extend the best-performing fusion method described in \cite{Watcharasupat2022AutonomousGain} to multiple modalities, such that in \textsc{ef}, we have
\vspace{-1mm}
\begin{align}
	f_{\text{g}}^{(\textsc{ef})}\left(
		\bm{k}, \bm{q}, \gamma, \bm{h}, \bm{r}
	\right) 
	&=
	\operatorname{Dense}\left(
		\operatorname{Conv}\left(
			\operatorname{Stk}\left(
				\bm{k},
				\bm{q},
				\bm{\Gamma},
				\mathbf{H},
				\mathbf{R}
			\right)
		\right)
	\right),
	\label{eq:f_g_EF}
\end{align}
\vspace{-4mm}

\noindent where $\bm{\Gamma} = \gamma\bm{1}_{N\times D}$,
$\mathbf{H} = \bm{1}_{N\times 1} \bm{h}^{\mathsf{T}}$, $\mathbf{R} = \bm{1}_{N\times 1} \bm{r}^{\mathsf{T}}$, $\bm{1}_{N\times D}$ is an $N$-by-$D$ matrix of ones, $\operatorname{Stk}\colon ( \mathbb{R}^{N\times D})^{B} \mapsto \mathbb{R}^{N\times D\times B}$ denotes ``channel-wise'' tensor stacking of all $B$ input arguments, $\operatorname{Dense}\colon\mathbb{R}^{N\times\bullet}\mapsto \mathbb{R}^{N\times D}$ is a dense layer with $D$ output units, and $\operatorname{Conv}\colon\mathbb{R}^{N\times D\times B}\mapsto\mathbb{R}^{N\times D}$ is a convolutional layer with the one-dimensional kernels compressing the stacked dimension into a singleton axis, thereby summarizing information from all modalities when performing soundscape augmentation in the feature domain. This definition of $f_{\text{g}}^{(\textsc{ef})}$ also prevents the problem of asynchronization of heterogeneous features mentioned in \cite{Chen2015Multi-modalNetworks}, since the ``synchronization'' occurs on the newly-created ``channel'' axis.

\vspace{-2mm}

\subsubsection{Mid-level fusion \textup{(\textsc{mf})}}\label{sec:Proposed Method/Contextual PPAP/Mid-level fusion}

In \textsc{mf}, the output block $f_{\text{o}}$ is modified to fuse information from all modalities such that they are included just before the output distribution $\mathcal{N}(\widehat{\mu},\widehat{\sigma}^2)$ is predicted. The feature augmentation block $f_{\text{g}}$ no longer uses $\bm{h}$ and $\bm{r}$ as inputs in \textsc{mf}. In other words, for \textsc{mf}, we have
\vspace{-4mm}
\begin{align}
	\begin{bmatrix}
		\widehat{\mu} & \log\widehat{\sigma}
	\end{bmatrix}^{\mathsf{T}}
	&=
	f_{\text{o}}^{(\textsc{mf})} \left(
		\operatorname{Concat}(\bm{z},\bm{h},\bm{r})
	\right), \text{ and}
	\label{eq:f_o_MF}
	\\
	f_{\text{g}}^{(\textsc{mf})}\left(
		\bm{k}, \bm{q}, \gamma
	\right) 
	&=
	\operatorname{Dense}\left(
		\operatorname{Conv}\left(
			\operatorname{Stk}\left(
				\bm{k},
				\bm{q},
				\bm{\Gamma}
			\right)
		\right)
	\right),
	\label{eq:f_g_MF}
\end{align}
\vspace{-4mm}

\noindent where $\operatorname{Concat}(\cdot)$ denotes concatenation along the embedding axis.

\vspace{-2mm}

\subsubsection{Late fusion \textup{(\textsc{lf})}}\label{sec:Proposed Method/Contextual PPAP/Late fusion}

In \textsc{lf}, an output adapter $A_{\text{o}}$ is added to perform fusion \textit{after} the final output of the \textsc{aPPAP}, such that the output distribution $\mathcal{N}(\widehat{\mu},\widehat{\sigma}^2)$ from the audio modality is transformed into $\mathcal{N}(\widehat{\mu}',(\widehat{\sigma}')^2)$ using all modalities. The predicted distribution is now $\mathcal{N}(\widehat{\mu}',(\widehat{\sigma}')^2)$, and
\vspace{-1.5mm}
\begin{align}
	\begin{bmatrix}
		\widehat{\mu}' & \log\widehat{\sigma}'
	\end{bmatrix}^{\mathsf{T}}
	&=
	A_{\text{o}} \left(
	\operatorname{Concat}(\widehat{\mu},\log\widehat{\sigma},
\bm{h},		\bm{r}
	)\right).
	\label{eq:f_o_LF}
\end{align}
\vspace{-4.5mm}

\noindent As with \textsc{mf}, the feature augmentation block $f_{\text{g}}$ does not use $\bm{h}$ and $\bm{r}$ as inputs in \textsc{lf}, so $f_{\text{g}}^{(\textsc{lf})} \equiv f_{\text{g}}^{(\textsc{mf})}$.

\vspace{-1mm}

\section{Validation Experiments}\label{sec:Validation Experiments}

We compared the mean squared error (MSE) of a \textsc{cPPAP} using the three fusion methods in \Cref{sec:Proposed Method/Contextual PPAP} in predicting the \textit{normalized ISO Pleasantness} (\textsc{isoPl}) of an augmented soundscape, with \textsc{ef}/\textsc{mf} and \textsc{lf} variants respectively predicting $\widetilde{\mu}_k = \widehat{\mu}_k$ and $\widetilde{\mu}_k = \widehat{\mu}'_k$ to obtain
\vspace{-2mm}
\begin{align}
	\textup{MSE} &= \frac{1}{K}\sum_{k}\left(y_k-\widetilde{\mu}_k\right). \label{eq:MSE}
\end{align}
\vspace{-4mm}

\noindent The \textsc{isoPl} is defined in \cite{Ooi2022ProbablyAugmentation} as a value in $[-1,1]$. As a further ablation study, we investigated the MSE for each combination of including/excluding participant and/or visual information.

For ease of reference, we denote variants with participant embeddings $\bm{h}$ included/excluded as \textsc{ip}/\textsc{ep}, and with visual embeddings $\bm{r}$ included/excluded as \textsc{iv}/\textsc{ev}. The baseline model for comparison was the \textsc{aPPAP}, corresponding to the \textsc{ep}+\textsc{ev} case, with the best-performing setup from \cite{Watcharasupat2022AutonomousGain}. This setup is detailed in \Cref{sec:Validation Experiments/Model architecture and training}. 

\vspace{-2mm}

\subsection{Dataset}\label{sec:Validation Experiments/Dataset}

We used the ARAUS dataset \cite{Ooi2022ARAUS:Soundscapes}, which contains a 5-fold cross-validation set of \num{25440} unique perceptual responses to augmented urban soundscapes presented as audio-visual stimuli.
\newpage
Corresponding information on the participants rating the stimuli was also collected via a participant information questionnaire (PIQ), consisting of basic demographic information and standard psychological questionnaires.
The \textsc{isoPl} values can be computed from each unique response and were used as the target observations for our validation experiments. The base soundscapes $\bm{s}$ and accompanying images $\bm{b}$ in the ARAUS dataset were drawn from the Urban Soundscapes of the World database \cite{DeCoensel2017UrbanMind}, with images extracted from the \SI{0}{\degree}-azimuth, \SI{0}{\degree}-elevation field of view (FoV) of the 30-second video captured at the same time as the 30-second soundscape recordings. When presented to the participants, the audio-visual stimuli in the ARAUS dataset used the entirety of the 30-second video at the same FoV, synchronized to the audio. For this study, we take a random frame from the 30-second video and downsample the frame via bilinear interpolation to obtain a standard image dimension of $(H,W,C_{\text{v}}) = (240,135,3)$ to be used as raw visual input to the \textsc{cPPAP}. This corresponds to a video frame rate of $\frac{\text{1}}{\text{30}}$ \SI{}{\hertz}. More frames, or the entirety of the video, could be used to extract a time series of visual embeddings corresponding to a higher frame rate and temporal relations between them could be explored in future work.

The 30-second maskers $\bm{m}$ were drawn from the Freesound and xeno-canto repositories, and calibrated as in \cite{Ooi2021AutomationHead} to obtain accurate log-gain values $\gamma$ if non-silent. If the masker was silent, the information in $\gamma$ was irrelevant, so we drew $\gamma$ from $\mathcal{N}(\nu,\zeta^2)$, where $\nu$ and $\zeta$ are the mean and standard deviation of the log-gains of the training set samples with non-silent maskers. This prevented the trained models from varying predictions at inference time according to $\gamma$ despite the soundscape (and hence ground-truth label) staying constant regardless of the value of $\gamma$ when the masker was silent. 

To obtain the coded participant information $\bm{p}$, we normalized all PIQ responses in the ARAUS dataset to the range $[0,1]$ if they corresponded to continuous variables (e.g., age), and converted them to binary dummy variables in $\{0,1\}$ if they corresponded to unordered categorical variables (e.g., dwelling type).

As an initial study, only a subset of PIQ items was selected. This was done by using the normalized PIQ responses as additional predictor variables to the elastic net \textsc{isoPl} model in \cite{Ooi2022ARAUS:Soundscapes}. Only variables with regression coefficients significantly different from zero ($p < 0.05$) were selected for use in the \textsc{cPPAP}. There were $M=5$ such participant-linked variables: their highest education attained, whether their main residence was landed property; their satisfaction of the overall acoustic environment in Singapore; their score on a modified Weinsten Noise Sensitivity Scale \cite{Weinstein1978}; and their Positive Affect score on the Positive and Negative Affect Schedule \cite{Watson1988}.

\subsection{Model architecture and training}\label{sec:Validation Experiments/Model architecture and training}

For the \textsc{aPPAP}, the audio feature extractors $f_{\text{s}}$ and $f_{\text{m}}$ comprise 5 convolutional blocks, each with a 3-by-3 convolutional layer, batch normalization, dropout, swish activation, and 2-by-2 average pooling. The numbers of filters in each block are 16, 32, 48, 64, 64, respectively. The spectrogram parameters are $T=644$, $F=64$, and $C_{\text{s}} = 2$, so the audio feature extractors give embeddings with dimension $N=20$ and $D=128$. The attention block $f_{\text{a}}$ uses dot-product attention \cite{Luong2015EffectiveTranslation} and the output block $f_{\text{o}}$ consists of 3 dense layers in sequence, with the first two having 128 units and swish activation, and the last having 2 units and linear activation.

For the \textsc{cPPAP}, the visual feature extractor $f_{\text{v}}$ has the same 5 convolutional blocks as $f_{\text{s}}$ and $f_{\text{m}}$, but pooling is performed using square grids of width 2, 2, 2, 3, and 5, respectively, such that the visual embeddings also have dimension $D = 128$. The participant feature extractor $f_{\text{p}}$ comprises a single dense layer with 128 units and swish activation. The output adapter $A_\text{o}$ consists of 3 dense layers in sequence, with the first two having $2^{\lfloor \log_2(M) \rfloor + 1} = 8$ units and swish activation, and the last having 2 units and linear activation.

All model types in the validation experiments were trained under a 5-fold cross-validation scheme with the same 10 seeds for each validation fold, for a total of 50 runs per model type. Each model was trained for up to 100 epochs using an Adam optimizer with a learning rate of \num{e-4}. In the \textsc{ep} and \textsc{ev} scenarios for the ablation study, we set $\bm{h}$ and $\bm{r}$ respectively to zero vectors. For the \textsc{aPPAP} (\textsc{ep}+\textsc{ev} scenario), both $\bm{h}$ \textit{and} $\bm{r}$ are set to zero vectors.

\section{Results and Discussion}\label{sec:Results and Discussion}

\Cref{tab:ablation-results} displays the results of the validation experiments and abalation study described in \Cref{sec:Validation Experiments}. All models performed better than the baseline \textsc{aPPAP} except for the \textsc{ip}+\textsc{iv} variants with mid-level and late fusion. To quantify the significance of any performance differences, we performed Kruskal-Wallis tests with Bonferroni correction between the \textsc{aPPAP} and the variants of the \textsc{cPPAP} investigated for this study. All \textsc{ip}+\textsc{ev} variants had significantly improved performance over the \textsc{aPPAP}, indicating that the fusion of additional information from the participant modality allowed the trained models to better predict the \textsc{isoPl} values derived from the ARAUS dataset responses. With late fusion, the \textsc{ip}+\textsc{ev} variant also performed the best among all investigated models with a cross-validation MSE of 0.1183$\pm$0.0011, a 2.8\% improvement over the \textsc{aPPAP}.

\begin{table}[t!]

\centering
    \begin{tabularx}{\columnwidth}{%
        >{\centering}X
        >{\centering}X
        >{\centering}X
        S[table-format=1.4]
            <{{ $\pm$ }}
            @{\hphantom{ $\pm$ }}
            S[table-format=1.4] 
        S[table-format=-1.1, 
            retain-explicit-plus=true]
        S[table-format=1.2,
            round-mode=places, 
            round-precision=3]
        l
    }
    \toprule
    \textbf{PIQ} &
    \textbf{Visual} &
    \textbf{Fusion} &
    \multicolumn{2}{c}{\textbf{MSE}} &
    \multicolumn{1}{c}{\textbf{$\%\Delta$}} &
    \multicolumn{2}{c}{\textbf{Adj. \textit{p}-val.}} \\
    \midrule
    \textsc{ep} & \hspace{2mm}\textsc{ev}           & \hspace{2.5mm}---         & 0.1217 & 0.0009 & 0.0 & {---} \\
    \midrule                                                          
    \textsc{ip} & \hspace{2mm}\textsc{ev}           & \hspace{2.5mm}\textsc{ef} & 0.1199 & 0.0009 & +1.5 & 0.01036840588327961332 & * \\
    \textsc{ip} & \hspace{2mm}\textsc{ev}           & \hspace{2.4mm}\textsc{mf} & 0.1194 & 0.0010 & +1.9 & 0.00342952606845617781 & * \\
    \textsc{ip} & \hspace{2mm}\textsc{ev}           & \hspace{2.4mm}\textsc{lf} & 0.1183 & 0.0011 & +2.8 & 0.00190964584100321169 & * \\
    \midrule                                                          
    \textsc{ep} & \hspace{2mm}\textsc{iv}           & \hspace{2.4mm}\textsc{ef} & 0.1204 & 0.0010 & +1.1 & 0.11350029689382520881 &   \\
    \textsc{ep} & \hspace{2mm}\textsc{iv}           & \hspace{2.4mm}\textsc{mf} & 0.1211 & 0.0011 & +0.5 & 0.86673322825819387738 &   \\
    \textsc{ep} & \hspace{2mm}\textsc{iv}           & \hspace{2.4mm}\textsc{lf} & 0.1204 & 0.0011 & +1.1 & 0.11350029689382520881 &   \\
    \midrule                                                          
    \textsc{ip} & \hspace{2mm}\textsc{iv}           & \hspace{2.4mm}\textsc{ef} & 0.1194 & 0.0012 & +1.9 & 0.00792668871667554821 & * \\
    \textsc{ip} & \hspace{2mm}\textsc{iv}           & \hspace{2.4mm}\textsc{mf} & 0.1250 & 0.0017 & -2.7 & 0.01348986003436508131 & * \\
    \textsc{ip} & \hspace{2mm}\textsc{iv}           & \hspace{2.4mm}\textsc{lf} & 0.1218 & 0.0016 & -0.1 & 1.00000000000000000000 &   \\
    \bottomrule
    \end{tabularx}

	\caption{Mean fold MSEs ($\pm$ standard deviation) over 10 unique seeds for the \textsc{isoPl} of the tested model configurations.
	Asterisks (*) denote statistically significant differences (adjusted-$p<0.05$) against the baseline audio-only PPAP (the \textsc{ep}+\textsc{ev} configuration).}
	\label{tab:ablation-results}
\end{table}

Moreover, the \textsc{ep}+\textsc{iv} variants, which used additional information from the visual modality, also performed better than the \textsc{aPPAP} but improvements were insignificant. This could be because the images used to derive the visual embeddings $\boldsymbol{r}$ were an \textit{objective} characteristic of the environment, whereas the participant embeddings $\boldsymbol{h}$ captured the participants' \textit{subjective} perception of the environment, thus making the \textsc{ep}+\textsc{iv} variants perform worse than the \textsc{ip}+\textsc{ev} variants. Alternative inputs from the visual modality better representing subjective perception could involve the subjectively-rated visual amenity and visual pleasantness \cite{Ricciardi2015SoundData}, if further data collection beyond the present iteration of the ARAUS dataset can be performed.

In addition, the \textsc{ip}+\textsc{iv} variants, which used information from both the participant and visual modalities, only had significant improvement over the \textsc{aPPAP} with early fusion at the feature augmentation block $f_g$. With mid-level and late fusion, the \textsc{ip}+\textsc{iv} variant actually performed worse than the \textsc{aPPAP}, which possibly hints at overfitting for the \textsc{mf} and \textsc{lf} scenarios due to the combined increase in number of non-acoustic predictor variables used at those stages. Therefore, early fusion via \Cref{eq:f_g_EF} is likely to be more suitable for the combination of multiple modalities.

Finally, at inference time, models using information from the participant modality can be used to simulate the ratings of hypothetical participants experiencing the same soundscape by varying $\boldsymbol{p}$ while keeping $\boldsymbol{s}$, $\boldsymbol{m}$, $\gamma$, and $\boldsymbol{b}$ constant. Averaging the ratings over a large variety of soundscapes, such as that in the ARAUS dataset, thus alllows us to isolate changes in perception due purely to participant-linked information while heightening the generalizability of both the models and the dataset.

\begin{figure}[!t]
    \centering
    \includegraphics[width=0.95\columnwidth]{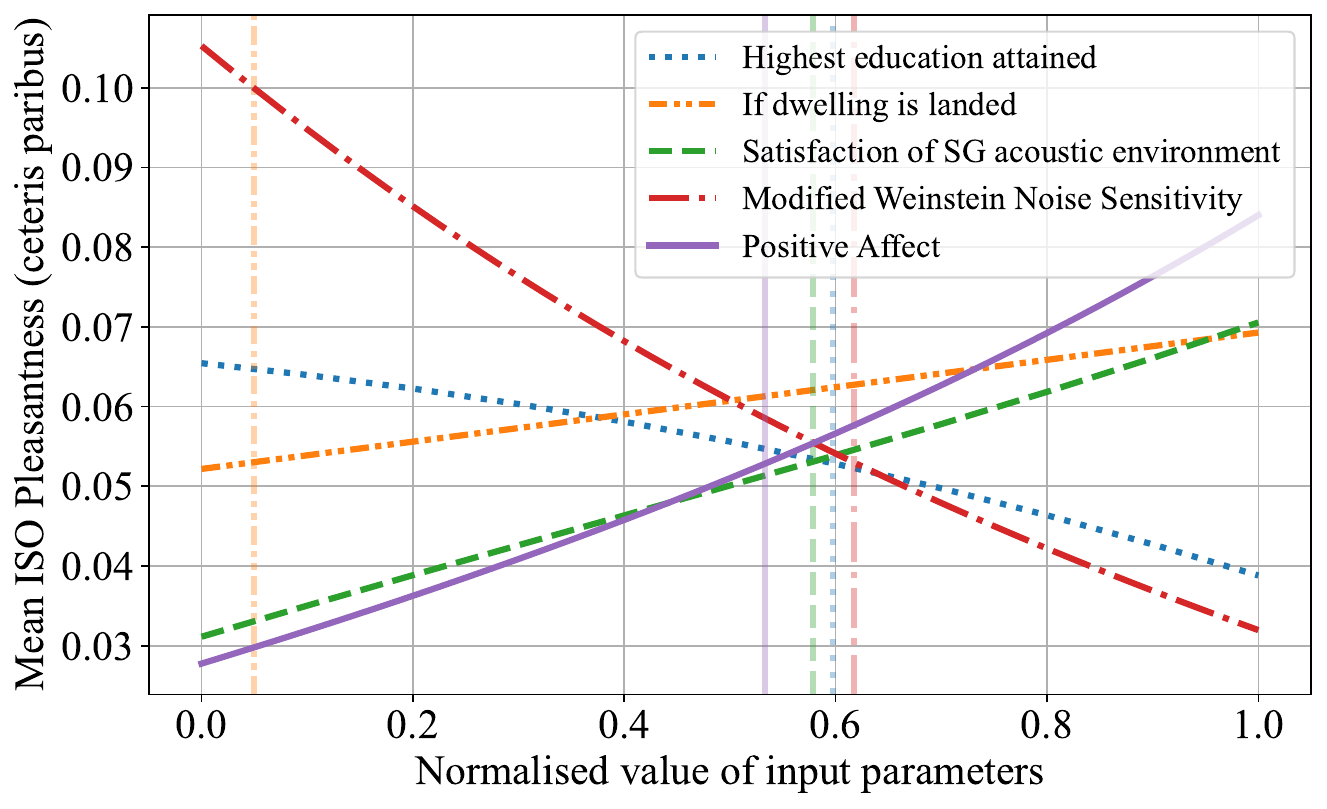}
    \vspace{-15pt}
    \caption{Mean \textsc{isoPl} predictions by the \textsc{cPPAP} (\textsc{ip}+\textsc{iv}+\textsc{ef} variant, seed 2) across all ARAUS dataset samples as a function of $[0,1]$-normalized PIQ items used in \Cref{sec:Validation Experiments}. Faded vertical lines denote the mean values of the same PIQ items within the ARAUS dataset.}
    \label{fig:demo-exploration}
    \vspace{-10pt}
\end{figure}

As an illustration, using a trained \textsc{ip}+\textsc{iv} model undergoing early fusion, we stimulated hypothetical participants experiencing all the audio-visual stimuli in the ARAUS dataset, and present the mean \textsc{isoPl} ratings in \Cref{fig:demo-exploration}. These participants were each represented by a different value of $\boldsymbol{p}$, where individual dimensions were varied while maintaining all other dimensions at their mean values in the training set (a ``ceteris paribus'' assumption). We can see, for instance, that mean \textsc{isoPl} ratings decrease nonlinearly with increasing noise sensitivity, and that there is a fairly linear relationship between the satisfaction that a hypothetical participant has with the overall acoustic environment in Singapore with the same \textsc{isoPl} ratings.

\section{Conclusion}\label{sec:Conclusion}

In conclusion, we proposed an architecture for a contextual PPAP that allows it to utilize multimodal features from the acoustic, visual, and participant domains in predicting perceptual ratings of augmented soundscapes while being compatible with its audio-only version via the zeroing of information from other domains. We established the efficacy of the modified architecture as \textit{ISO Pleasantness} models trained using the ARAUS dataset, and demonstrated how the contextual PPAP could also be used as a model to observe the impact of demographic factors on soundscape perception. Future work could involve the deployment of the contextual PPAP as a pre-trained model in an automatic masker selection system, followed by in-situ verification experiments to assess the ecological validity of the results obtained in this study in a real-life deployment context. Alternatively, the impact of other contextual factors, such as physiological measurements and ambient weather conditions, can be explored as additional input modalities for the contextual PPAP.

\FloatBarrier

\bibliographystyle{IEEEbib}

\end{document}